\documentclass[fleqn,usenatbib]{mnras}

\usepackage{mathptmx}
\usepackage{txfonts}

\usepackage[T1]{fontenc}
\usepackage{ae,aecompl}

\usepackage{graphicx}	
\usepackage{amssymb}	

\usepackage{array}
\usepackage{lipsum}
\usepackage{textcomp}

\newcolumntype{X}{>{$}c<{$}} 
\newcolumntype{Z}{>{$}l<{$}} 
\newcolumntype{C}{>{$}r<{$}} 

\newcommand{\Msun}{\mbox{$\mathrm{M}_{\odot}$}}
\newcommand{\Teff}{\mbox{$T_{\mathrm{eff}}$}}
\newcommand{\logg}{\mbox{$\log g$}}
\newcommand{\hhe}{\mbox{$\log\mathrm{[H/He]}$}}
\newcommand{\Rsun}{\mbox{$\mathrm{R}_{\odot}$}}

\newcommand{\dpi}{\mbox{$d_\varpi$}}

\newcommand{\Ion}[2]{#1{\,\textsc{#2}}}

\newcommand{\Maccrate}{8.8\,\times\,10^{8}\mathrm{\,g\,s^{-1}}}
\newcommand{\wdjname}{WD\,J204713.76--125908.9}
\newcommand{\wdj}{WD\,J2047--1259}

\usepackage[usenames,dvipsnames]{xcolor}
\usepackage{soul}

\definecolor{rr}{HTML}{2ecc71}

\title[Accretion onto \wdjname]{White dwarf pollution by hydrated planetary remnants: Hydrogen and Metals in \wdjname}

\author[M. J. Hoskin et al.]{Matthew J. Hoskin,$^{1,2}$\thanks{E-mail: M.Hoskin@warwick.ac.uk}
Odette Toloza,$^{1}$ Boris T. G\"ansicke,$^{1,2}$ Roberto Raddi,$^{3,4}$
\newauthor
Detlev Koester,$^{5}$ Anna F. Pala,$^{6,1}$ Christopher J. Manser,$^{1}$ Jay Farihi,$^{7}$
\newauthor
Maria Teresa Belmonte,$^{8}$ Mark Hollands,$^{1}$ Nicola Gentile Fusillo,$^{6,1}$ Andrew Swan$^{7}$
\\
$^{1}$Department of Physics, University of Warwick, Coventry, CV4 7AL, UK\\
$^{2}$Centre for Exoplanets and Habitability, University of Warwick, Coventry, CV4 7AL, UK\\
$^{3}$Dr. Remeis-Sternwarte \& ECAP, Friedrich-Alexander Universit\"at Erlangen-N\"urnberg, Sternwartstr. 7, 96049 Bamberg, Germany\\
$^{4}$Departament de F\`isica, Universitat Polit\'ecnica de Catalunya, c/Esteve Terrades 5, E-08860 Castelldefels, Spain\\
$^{5}$Institut f\"ur Theoretische Physik und Astrophysik, Christian-Albrechts-Universit\"at, 24118 Kiel, Germany\\
$^{6}$European Southern Observatory, Karl Schwarzschild Stra{\ss}e 2, Garching, 85748, Germany\\
$^{7}$University College London, London, WC1E 6BT, UK\\
$^{8}$Blackett Laboratory, Physics Department, Imperial College London, London SW7 2AZ, UK
}
\date{Accepted XXX. Received YYY; in original form ZZZ}
\pubyear{2019}

\begin{document}
\label{firstpage}
\pagerange{\pageref{firstpage}--\pageref{lastpage}}

\maketitle

\begin{abstract}
\wdjname\ is a new addition to the small class of white dwarfs with  helium-dominated photospheres that exhibit strong Balmer absorption lines and atmospheric metal pollution.
The exceptional abundances of hydrogen observed in these stars may be the result of accretion of water-rich rocky bodies.
We obtained far-ultraviolet and optical spectroscopy of \wdjname\ using the Cosmic Origin Spectrograph on-board the \textit{Hubble Space Telescope} and X-shooter on the Very Large Telescope, and identify photospheric absorption lines of nine metals: C, O, Mg, Si, P, S, Ca, Fe and Ni.
The abundance ratios are consistent with the steady state accretion of exo-planetesimal debris rich in the volatile elements carbon and oxygen, and the transitional element sulphur, by factors of seventeen, two, and four respectively compared to bulk Earth.
The parent body has a composition akin to Solar System carbonaceous chondrites, and the inferred \emph{minimum} mass, $1.6 \times 10^{20}$\,g, is comparable to an asteroid $23$\,km in radius.
We model the composition of the disrupted parent body, finding from our simulations a median water mass fraction of eight~per~cent.

\end{abstract}



\section{Introduction}
 
White dwarfs are remnants of main sequence stars less massive than $\simeq8$\Msun, which have shed their envelope during their post-main sequence evolution to leave a cooling degenerate star.
Their high surface gravities ($\logg\simeq8$) cause chemical stratification, so their atmospheres are dominated by either hydrogen or helium.
Studies of increasing sensitivity have shown that the majority of helium atmosphere white dwarfs below $\simeq20,000$\,K contain detectable atmospheric hydrogen, but insufficiently abundant to form a superficial radiative hydrogen layer \citep[]{Fontaine-1987-conf-SpecEv, Eisenstein-2006-AJ-DBTrough, Voss-2007-AA-SN1aDB, Bergeron-2011-ApJ-DB_spec, Koester-2015-AA-DBSurvey, Rolland-2018-ApJ-DBAs}.
Spectra of these stars can show absorption lines from both elements, and their relative abundance, \hhe\footnote{$\hhe\,=\,\log [\mathrm{n(H)}/\mathrm{n(He)}]$, where n is numerical abundance.}, can be determined from model atmosphere analysis. In most cases where both elements are present, helium is the dominant element, with \hhe\ ranging from $10^{-1}$ to $10^{-7}$ or lower.

Spectral evolution of white dwarfs between hydrogen- and helium-dominated atmospheres has been studied extensively; however, the origin of hydrogen remains contentious \citep[e.g.][]{Koester-1976-AA-ConvectiveMixing, Fontaine-1987-conf-SpecEv, MacDonald-1991-ApJ-WD_hydrogen, Beauchamp-1996-conf-DBAs,Voss-2007-AA-SN1aDB, Tremblay-2008-ApJ-DB_fraction, Koester-2015-AA-DBSurvey, GentileFusillo-2017-MNRAS-He_water, Rolland-2018-ApJ-DBAs, Cunningham-2020-MNRAS-SpecEvol}.
Throughout the literature, trace hydrogen in helium atmospheres has been ascribed to:
(i) convective processes dispersing hydrogen that survived giant branch evolution \citep{Herwig-1999-AA-HDeficient, Althaus-2010-AARv-WDEvolution, Koester-2015-AA-DBSurvey, Rolland-2018-ApJ-DBAs, Rolland-2020-arXiv-HDredge};
(ii) episodic or continuous accretion from the interstellar medium \citep[ISM:][]{Koester-1976-AA-ConvectiveMixing, MacDonald-1991-ApJ-WD_hydrogen, Voss-2007-AA-SN1aDB};
(iii) water-rich planetesimal accretion \citep{Farihi-2013-Science-Water_evidence, Veras-2014-MNRAS-Oort, Raddi-2015-MNRAS-WaterRich, GentileFusillo-2017-MNRAS-He_water, Xu-2017-ApJl-KBO}.

\citet[]{Rolland-2018-ApJ-DBAs, Rolland-2020-arXiv-HDredge} showed that many measured \hhe\ values in their sample are incompatible with evolution involving either a constant mass of (primordial) hydrogen, or a constant rate of hydrogen accretion, since the white dwarf formed.
However, along with the presence of trace hydrogen, trace metals are commonly seen in the atmospheres of helium (and hydrogen) atmospheres white dwarfs \citep[e.g.][]{Koester-2015-AA-DBSurvey, GentileFusillo-2017-MNRAS-He_water}.
Unlike hydrogen, which persists in the upper atmosphere, metals sink below the photosphere on timescales between $10^{-2}$--$10^{7}$\,yr \citep{Koester-2009-AA-NewAccDiff}.
The consensus is that these metals originate in the star's planetary system; from rocky bodies which have been captured and accreted.

Planetesimals perturbed onto orbits that cross the Roche radius of the white dwarf will undergo tidal disruption and form a debris disc, before being accreted \citep{Jura-2003-ApJl-DisruptedAsteroid, Jura-2008-AJ-Asteroid_pollution, Metzger-2012-MNRAS-RunawayAcc, Veras-2014-MNRAS-TidalDisruption, Veras-2015-MNRAS-ShrinkingRings}.
Debris discs have been detected at infrared wavelengths as an excess of flux in tens of systems \citep[e.g.][]{Graham-1990-ApJ-G29-38, Kilic-2007-ApJ-DustyDisk, vonHippel-2007-ApJ-SpitzerSurvey, Debes-2011-ApJS-IRExcesses, Girven-2012-ApJ-DiskLifetimes, Brinkworth-2012-ApJ-4Disks, Rocchetto-2015-MNRAS-IRDiscs, Farihi-2016-NAR-CircumstellarDebris}, and through emission lines from gaseous metals in a few cases \citep[e.g.][]{Gaensicke-2006-Science-GasDisk, Gaensicke-2007-MNRAS-GasDisk2, Gaensicke-2008-MNRAS-GasDisk0845, Farihi-2012-MNRAS-GasDiskTrio, Melis-2012-ApJL-GasDiskHE1349}.

Once accreted, metals give rise to narrow absorption lines which are detectable in spectroscopy \citep{Jura-2014-AREPS-Cosmochemistry, Hollands-2017-MNRAS-CoolDZ1}.
The bulk composition of the disrupted parent body can be obtained by modelling the atmospheric abundances to reproduce these spectroscopic lines, and the calculation of gravitational diffusion timescales for each element.
For atmospheres with effective temperature $\Teff \gtrsim 25\,000$\,K, radiative levitation can sustain metals at detectable abundances and so must also be considered \citep{Chayer-1995-AJ-RadLev}.
The variety in deduced compositions is comparable to our own Solar System \citep[e.g.][]{Koester-2005-AA-HS0146,Zuckerman-2007-ApJ-G362,Jura-2008-AJ-Asteroid_pollution,Melis-2011-ApJ-Accretion_minorplanet,Gaensicke-2012-MNRAS-PollutionDiversity,Xu-2013-ApJ-PlanetesimalEvolution}, and recent works have shown that this diversity can be explored quantitatively in terms of planet formation theory \citep{Harrison-2018-MNRAS-PollutedWDs, Swan-2019-MNRAS-Diversity} and Solar System chemistry \citep{Doyle-2019-Science-OFugacity}.

The possibility of detecting the accretion of water-rich planetary material was first raised by \citet[]{Jura-2009-ApJ-TwoPollutedWDs}.
More recent and extensive modelling~--~including factors such as formation, stellar progenitor mass and treatment of both heat and mass transport \citep{Malamud-2016-ApJ-PMSIce1, Malamud-2017-ApJ-PMSIce2, Malamud-2017-ApJ-PMSIce3}~--~has demonstrated that a large fraction of the total water content\footnote{Including both water ice and hydrated minerals~--~throughout this paper, we will simply refer to ``water" to refer to both.} of a rocky body can survive.
\citet{Farihi-2013-Science-Water_evidence} first reported a white dwarf with atmospheric metal abundances consistent with excess oxygen~--~indicative of water rich material~--~and evidence of similar accretion episodes has since been found \citep{Raddi-2015-MNRAS-WaterRich,Farihi-2016-MNRAS-WaterRich}.
Furthermore, eight of the 17 polluted white dwarfs examined by \citet{Harrison-2018-MNRAS-PollutedWDs} are best fitted by a parent body containing significant amounts of water, with five of these at $3\sigma$ statistical significance.
This demonstrates that some water sequestered in planetesimals survives into the later stages of stellar evolution, in line with the prediction of \citet{Jura-2010-AJ-WaterSurvival}.
Hydrogen atoms from dissociated water will remain in the white dwarf's atmosphere indefinitely, and is an observable indicator that water-rich material may have been accreted.

A statistical analysis of helium atmosphere white dwarfs found a highly significant correlation between photospheric metal pollution and the presence of trace hydrogen, interpreted as evidence that hydrogen is accreted alongside planetary debris~--~most likely as water \citet{GentileFusillo-2017-MNRAS-He_water}.
This is a compelling indicator that accretion of rocky planetary bodies can contribute to photospheric and spectral evolution of white dwarfs.
Further, it shows that in some cases hydrogen should be considered as an external pollutant alongside the metals often seen in white dwarf atmospheres.

There are seven known white dwarfs with a photosphere composed mainly of helium, but containing sufficient hydrogen that the Balmer lines dominate their spectral classification, and with existing observations that allow identification of atmospheric pollution:
GD\,16 \citep{Koester-2005-AA-HS0146}; GD\,362; PG\,1225--079 \citep[both][]{Xu-2013-ApJ-PlanetesimalEvolution}; SDSS\,J124231.06+522626.7 \citep{Raddi-2015-MNRAS-WaterRich}; WD\,1536+520 \citep{Farihi-2016-MNRAS-WaterRich}; WD\,1425+540 \citep{Xu-2017-ApJl-KBO}; GD\,17 \citep{GentileFusillo-2017-MNRAS-He_water}.
Multiple metals are identified in each of these stars, from which the compositions of accreted planetary material are measured.
This small family provides a valuable opportunity to probe the evolution of white dwarf atmospheres and investigate connections to planetesimal accretion.

In this paper we report on a new member of this family: \wdjname\ \citep[adopting the new white dwarf naming convention outlined by][hereafter referred to as \wdj]{GentileFusillo-2019-MNRAS-GDR2_WDs}.
This star was identified to have a hydrogen atmosphere \citep[as EC\,20444--1310]{O'Dononghue-2013-MNRAS-ECBOS3}, until \citet[as APASS\,J2047--1259]{Raddi-2017-MNRAS-AllSkySample} showed that Balmer and helium absorption lines are both present in optical spectra, and used $\hhe=-1$ to analyse the spectrum.
Neither \citet{O'Dononghue-2013-MNRAS-ECBOS3} nor \citet{Raddi-2017-MNRAS-AllSkySample} had observations with the required spectral coverage and resolution to probe for the presence of metals.
In Section~\ref{s:Obs}, we present follow-up spectroscopy taken with the \textit{Hubble Space Telescope} (\textit{HST}) and the Very Large Telescope (VLT), and in Section~\ref{s:Fitting}, we fit the atmospheric parameters of \wdj.
In Section~\ref{s:Comp}, we discuss the composition and nature of the accreted material, and in Section~\ref{s:IR} consider the infrared data and possibility of a dusty disc.
Section~\ref{s:HinHe} comprises further discussion of the spectral evolution of helium atmosphere white dwarfs and of \wdj, and in Section~\ref{s:Conc}, we summarise our findings and their implications.

\section{Observational Data}
\label{s:Obs}

\subsection{Spectroscopy}

Owing to the extraordinary hydrogen abundance measured by \citet{Raddi-2017-MNRAS-AllSkySample}, \wdj\ was included in the target list for our \textit{HST} snapshot program (ID:\,15073, Cycle 25) to search for photospheric pollution.
A short exposure ($2000$\,s) far-ultraviolet (FUV) spectrum was obtained with the Cosmic Origins Spectrograph (COS) on 2017 November 04 using the G130M grating, centred at $1291$\,\AA, with the instrument in Lifetime Position 4.
This set-up provides wavelength coverage of 1130~--~1431\,\AA\ (with a small gap at 1274~--~1287\,\AA) and resolving power $R\simeq16\,000$.

To mitigate the fixed pattern noise that is affecting the COS FUV detector, we divided the exposure time equally between two FP-POS positions (1 and 4; the limited snapshot visit duration did not allow us to use all four positions).
This spectrum reached a signal-to-noise of $15$, dominated by the broad Ly~$\alpha$ absorption feature.
A visual inspection identified strong absorption lines of silicon, carbon, oxygen, iron sulphur, phosphorus, and nickel.

\begin{figure*}
\includegraphics[width=\textwidth]{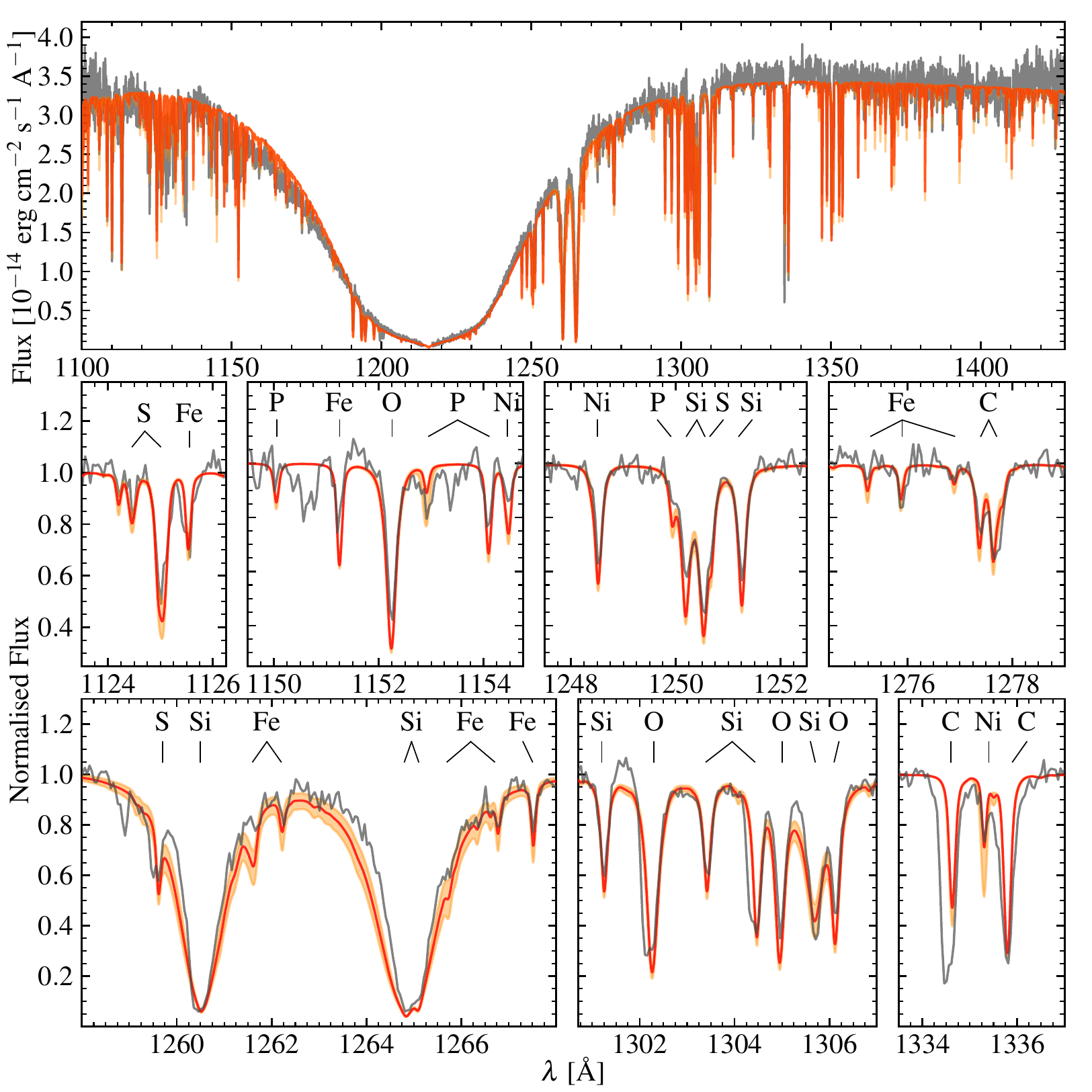}
\caption{Atmospheric model (red, with $\pm1\sigma$ uncertainty in orange) and the rebinned COS spectrum (grey). The top panel shows the quality of the overall fit, which notably deviates from the level of the continuum below 1190\,\AA\ (see main text). The middle and lower rows of panels show the strongest absorption lines of the atmospheric metals. The $1334.53$\r{A} \Ion{C}{ii} line is clearly blended with a significant ISM absorption line.
}
\label{f:MetalAbsorption}
\end{figure*}

To increase sensitivity~--~particularly in the Ly~$\alpha$ wings~--~and wavelength coverage, we carried out observations of a five-orbit mid-cycle GO program on 2018 November 18 (ID:\,15474, Cycle 26).
Two orbits used the G130M grating centred at $1291$\,\AA, in lifetime positions 3 and 4 (1 and 2 were unavailable).
The remaining three orbits used a central wavelength of $1222$\,\AA, and the exposure time was divided equally across all four lifetime positions.
At this central wavelength the Ly~$\alpha$ core and the airglow emission fall onto the detector segment gap ($1211$~--~$1222$\,\AA).
The wavelength coverage of this mode is $1064$~--~$1368$\,\AA, and the resolving power slightly lower at $R\simeq15\,000$.
\textit{HST} COS spectra were co-added and rebinned into a single spectrum with resolution elements of $0.05$\,\AA, achieving a signal-to-noise of 38 in the continuum.
This spectrum is shown in the upper panel of Fig.~\ref{f:MetalAbsorption}.
Time-tag data were visually inspected and show no variability.

\begin{figure*}
\includegraphics[width=\textwidth]{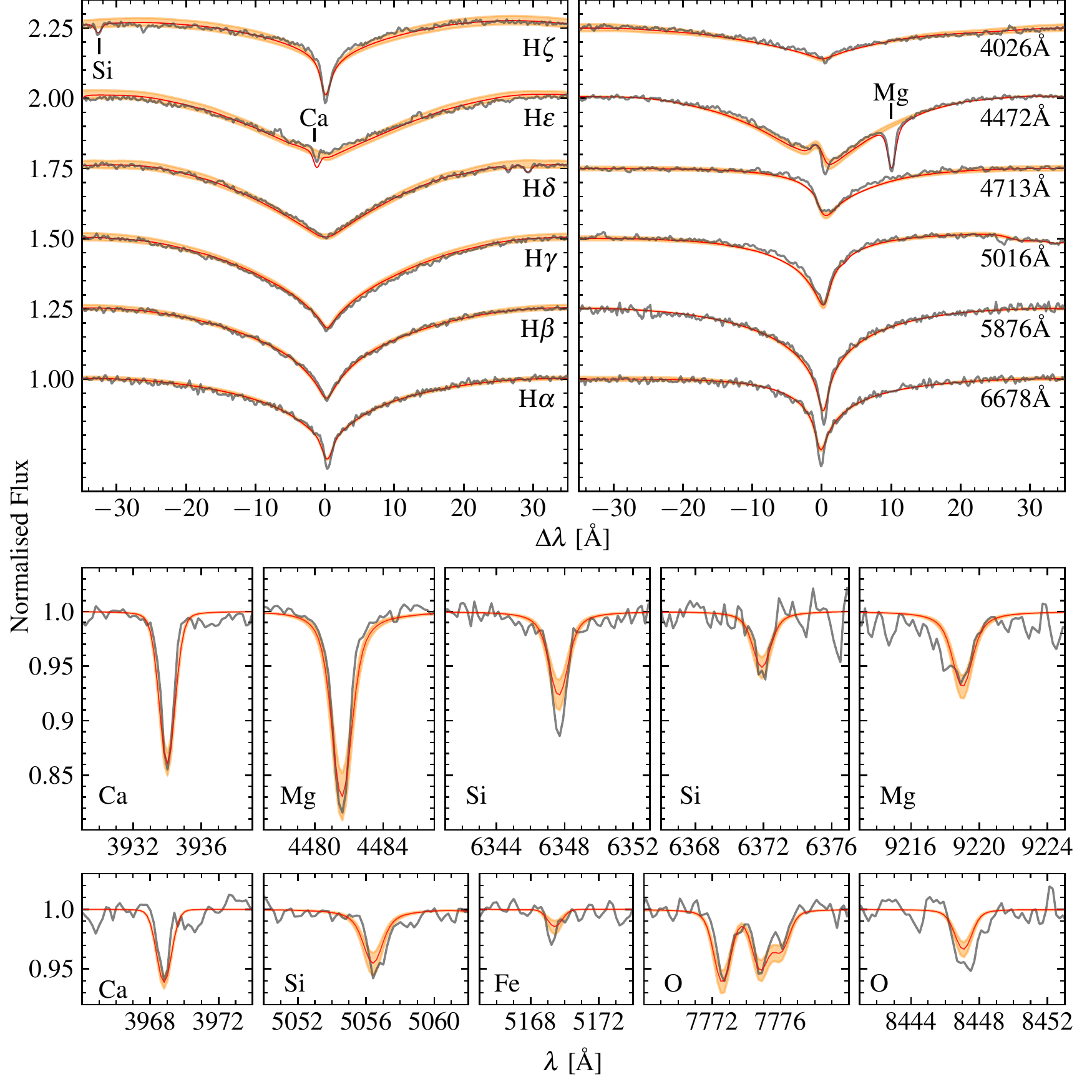}

\caption{Atmospheric model (red, with $\pm1\sigma$ uncertainty in orange) and the normalised X-shooter spectrum (grey).
The top panels show the normalised Balmer (left) and \Ion{He}{I} (right) features, labelled by name or by central wavelength, where the shaded region shows the uncertainty on the normalised model. Absorption lines from metal ions are also labelled. The middle and lower rows of panels highlight several strong metal absorption lines. In these panels the shaded region is associated with the uncertainty in metal abundances.
}
\label{f:MetalAbsorptionOptical}
\end{figure*}

On 2018 July 10, we obtained a deep optical spectrum using X-shooter \citep{Vernet-2011-AAP-XSHOOTER}, an echelle medium resolution spectrograph mounted on the Cassegrain focus of UT2 at European Southern Observatory's Very Large Telescope (ESO VLT) in Cerro Paranal (Chile).
We used slit widths of $1.0$, $0.9$, and $0.9$\,arcsec in the blue (UVB, $3000$~--~$5600$\,\AA, R$=5400$), visual (VIS, $5500$~--$10\,200$\,\AA, R$=8900$) and near-infrared (NIR, $10\,200$~--~$24\,800$\,\AA, R$=5600$) arms and exposed for $4 \times 1220$, $4 \times 1250$, and $4 \times 1300$\,s, giving median signal-to-noise ratios of 212, 126, and 53, respectively.
The NIR arm was read out using the non-destructive mode to mitigate the effect of overexposed sky lines.
Data were reduced using the \texttt{Reflex} pipeline \citep{reflex}, which includes the standard reduction steps of bias and dark current removal, order identification and tracing, flat-fielding, dispersion solution, correction for instrument response and atmospheric extinction, and merging of all orders. Flux calibration was achieved using observations of a standard star and the wavelength scales were corrected for the barycentric radial velocity. Finally, a telluric correction was performed using \texttt{molecfit} \citep{molecfit1,molecfit2}.
Absorption lines from calcium, magnesium, oxygen and iron, along with the broad Balmer and helium lines, are visible in the UVB and VIS arms, and are highlighted in Fig.~\ref{f:MetalAbsorptionOptical}.

\subsection{Photometry}

We have compiled the photometry of \wdj\ in Table~\ref{t:Photometry}.
Data were taken from the \textit{GALEX} \citep{Bianchi-2011-APSS-GALEX}, \textit{Gaia} \citep{Gaia-2016-AA-GaiaInstrument}, APASS \citep{Henden-2015-conf-APASS}, Pan-STARRS \citep{Chambers-2016-AAS-PanSTARRS}, 2MASS~\citep{Cutri-2003-conf-2MASS} and VHS \citep{Sutherland-2015-AA-VISTA} online catalogues.

\wdj\ was observed twice by the Spitzer Space Telescope \citep{Werner-2004-ApJ-Spitzer} during the post-cryogenic mission.
Observations in each warm IRAC channel were taken separately: at 3.6 $\mu$m on 2012 Jan 06, and at 4.5 $\mu$m on 2011 Jun 22 (both from program 80063, PI: Boyer).
The field was observed with 18 dithered frames using 30\,s exposures during each visit.
The target is well exposed and, owing to the small dithering scale, is well covered in the image stack at 90 arcsec from the edge of the mosaic.
The target lies 4.6 arcsec from an unrelated star with late M-dwarf colours based on \textit{Gaia} and 2MASS, and 5.0 arcsec from a background galaxy based on its diffuse IRAC image.

Mosaics with 0.6 arcsec pixel$^{-1}$ were created from the corrected basic calibrated data frames using the {\sc mopex} package, following standard observatory recommendations.
The flux was measured both using apertures and with the {\sc apex} module to perform point-spread function fitting simultaneously on the target and neighbouring sources.
Aperture and array-location-dependent corrections were made, but pixel-phase and color corrections were ignored.
The fluxes measured with an $r = 4$ pixel (2.4\,arcsec, 2 native pixel) radius aperture are likely contaminated by the neighbouring M star ($114 \pm 12$ and $66\pm7$\,$\mu$Jy for the $3.6\,\mu$m and $4.5\,\mu$m band passes respectively), while values reported by {\sc apex} appear consistent with the model predictions for the stellar photosphere ($67\pm4$ and $44\pm3$\,$\mu$Jy).
Photometric errors are the quadrature sum of the measurement error and a calibration uncertainty of 5 per~cent.

\begin{table}
\centering
\caption{Photometric data for \wdj.}
\label{t:Photometry}
\begin{tabular}{ p{24mm} Z X }
\hline
 Survey & \textnormal{Filter} & \textnormal{Brightness [AB mag]} \\
 \hline
 \textit{GALEX} & FUV & 15.531\pm0.012\\
 & NUV & 15.521\pm0.008\\
 APASS & B & 15.91\pm0.08 \\
 & g' & 15.86\pm0.03 \\
 & V & 15.89\pm0.06 \\
 & r' & 16.05\pm0.05 \\
 & i' & >15.97 \\
 Pan-STARRS & g & 15.901\pm0.003 \\
 & r & 16.180\pm0.002 \\
 & i & 16.466\pm0.003 \\
 & z & 16.718\pm0.005 \\
 & y & 16.875\pm0.009 \\
 \textit{Gaia} & BP & 15.965\pm0.009\,^a\\
 & G & 16.126\pm0.008\,^a\\
 & RP & 16.533\pm0.010\,^a\\
 2MASS & J & 17.35\pm0.16 \\ 
 & H & >16.40\\
 & K_s  & >16.49 \\
 VHS & Y & 16.903\pm0.008 \\
 & J & 17.318\pm0.013 \\
 & K & 18.34\pm0.06 \\
 \textit{Spitzer} & 3.6\,\mathrm{\mu}\mathrm{m} & 19.33\pm0.05 \\
 & 4.5\,\mathrm{\mu}\mathrm{m} & 19.79\pm0.05 \\
 \hline
\end{tabular}
\begin{flushleft}
$^a$\,Corrected using \citet{MaizApellaniz-AA-2018-GaiaDR2Corr}.
\end{flushleft}
\end{table}

\section{Spectroscopic fitting}
\label{s:Fitting}

We first undertook a rigorous fit to the atmospheric parameters, \Teff, \logg, and \hhe, using a bespoke model grid computed with the code of \citet{Koester-2010-MEMSAI-WDModels}.
Once these parameters were finalised, their values were fixed during a visual fitting of the atmospheric metals.

\subsection{Atmospheric Parameters}
\label{s:Atmos}

\begin{figure*}
\includegraphics[width=\textwidth]{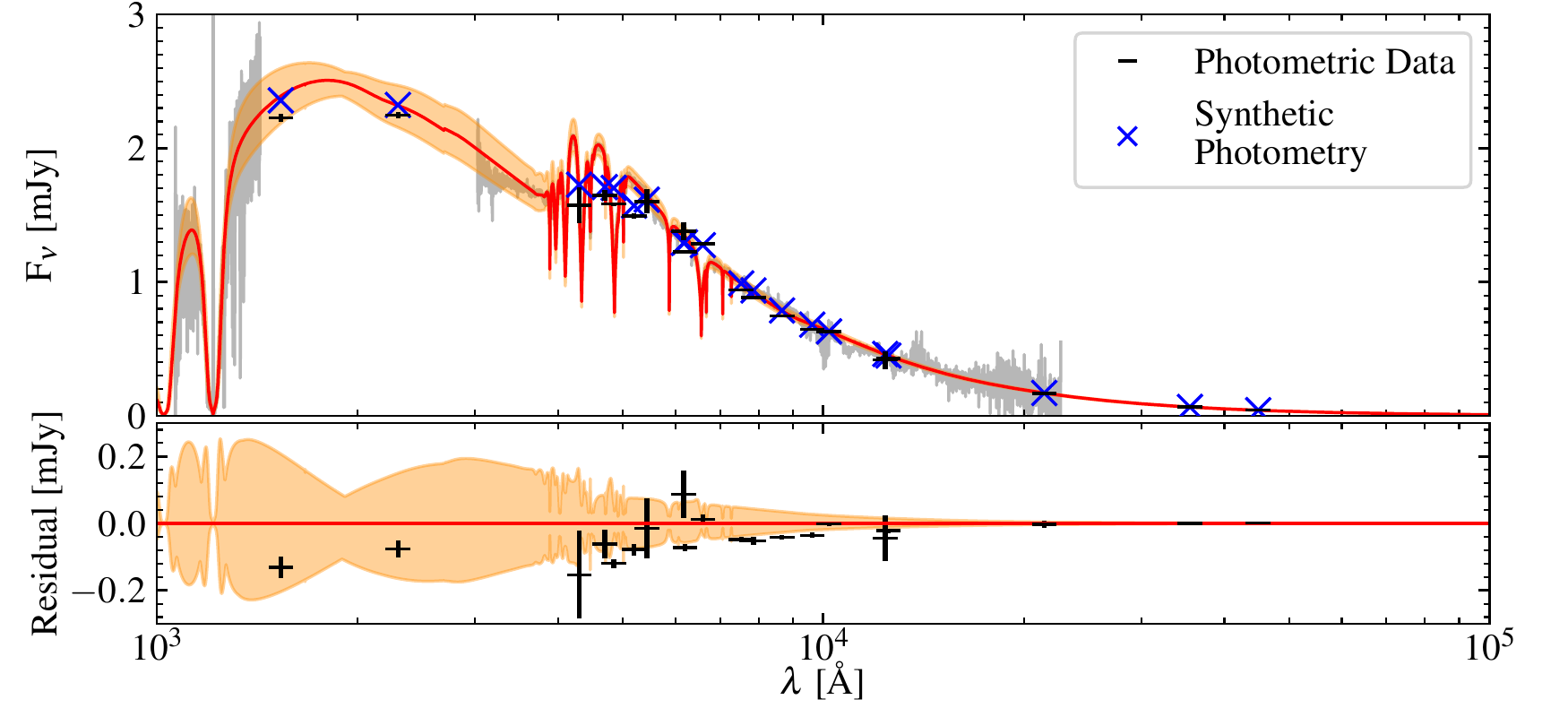}
\caption{Spectral energy distribution of \wdj.
From the model (red), which is the best-fit to the spectra (grey), synthetic photometry (blue crosses) are calculated and compared to survey photometry (black dashes).
Most of the photometric data lie below the best-fitting model, but most are close to or within the $1\sigma$ uncertainties in the spectroscopic fit.}
\label{f:SED}
\end{figure*}

Our grid of synthetic spectra covers $10\,000\le\Teff\le20\,000$\,K in steps of $200$\,K, $7.0\le\logg\le9.0$ in steps of $0.1$ dex, and $-5.0\le\hhe\le0.0$ in steps of $0.2$ dex, and uses ML2/$\alpha = 1.0$.
These models do not include metals.
We used the $\chi^2$ minimisation routine (\texttt{scipy.optimize.fmin}) from \citet{SciPy-2001} and tri-linearly interpolated synthetic spectra from our model grid.
For each synthetic spectrum, we computed the white dwarf radius, $R_\mathrm{WD}$, for the corresponding \Teff\ and \logg\ using the mass-radius relation\footnote{\url{http://www.astro.umontreal.ca/~bergeron/CoolingModels/}} of  \citet{Fontaine-2001-PASP-CosmoChronology}, and then applied a flux scaling factor of $S\,=\,R_\mathrm{WD}^2 / \dpi^2$, where \dpi\ is the \textit{Gaia} parallax distance.

\wdj\ has a parallax distance of just under $100$\,pc and so the effect of reddening is expected to be low.
Local 3D dust maps yield an extinction of $E(B\,-\,V)~=~0.005\pm0.017$\,mag \citep{Lallement-2014-AA-LocalDustMap, Capitanio-2017-AA-LocalDustMap, Lallement-2018-AA-LocalDustMap}.
We performed the fits described below using $E(B\,-\,V)~=~\{0.00,  0.01,  0.02,  0.03\}$\,mag and reddened the synthetic spectra using the global mean extinction law of \citet{Fitzpatrick-1999-PASP-Extinction}.
We found the photometric and spectroscopic fits were most consistent for $E(B\,-\,V)=0.01$, and we adopted this value for the final analysis.

We simultaneously fitted the COS spectrum (with metal absorption lines masked) and the normalised Balmer and helium line profiles in the X-shooter data\footnote{The near-infrared arm data are not used in our fitting procedure.} to obtain a self-consistent three-parameter (\Teff, \logg, \hhe) fit.
We draw 1000 data sets\footnote{We varied the number of samples over three orders of magnitude ($n = 100$ to $10\,000$), and found $n=1000$ gave results and distributions consistent with $n=10\,000$, so for computational expediency we used $n=1000$.} from a Monte Carlo simulation incorporating uncertainties from spectral extraction, and parallax for the un-normalised COS spectrum.
The scatter in \Teff, \logg, \hhe\ were used to estimate a Gaussian error on each parameter.
The best-fitting atmospheric parameters measured from the above procedure are  $\Teff=17\,970^{+140}_{-170}$, $\logg=8.04\pm0.05$ and $\hhe=-1.08\pm0.08$.
The spectroscopic model (including metals, discussed in Section~\ref{s:ZAbund}) and data are shown in Fig. \ref{f:MetalAbsorption} and Fig. \ref{f:MetalAbsorptionOptical}.
The Pearson correlation coefficient between \Teff\ and \logg\ is $\rho\,=\,0.99$, and between those and \hhe, $0.573$ and $0.596$ respectively.
These atmospheric parameters are broadly consistent with the result of the pre-\textit{Gaia} spectroscopic fit with $\hhe=-1$ fixed \citep{Raddi-2017-MNRAS-AllSkySample}.

We fitted models to the photometry in Table\,\ref{t:Photometry} with the \textit{Gaia}-based distance to check for consistency, but fix $\hhe=-1.08$ as broad-band photometry is poorly suited to constraining the hydrogen abundance.
We excluded \textit{Spitzer} data in case of an infrared excess from a debris disc, which we discuss in Section~\ref{s:IR}.
This recovers $\Teff\,=\,17\,920\pm60$\,K and $\logg\,=\,8.076\pm0.005$ with a Pearson correlation coefficient of $\rho\,=\,0.98$, which is consistent with our spectroscopic fit.
The purely statistical errors on the photometric data do not represent the true uncertainties in ground based photometry.
Similarly \textit{GALEX} and \textit{Gaia} photometry have been closely scrutinised, and significant corrections have been suggested for each \citep[respectively]{MaizApellaniz-AA-2018-GaiaDR2Corr, Wall-2019-MNRAS-GALEX-Corr}.
A comparison of the photometric data with the synthetic photometry from the \emph{spectroscopic} model is shown in Fig. \ref{f:SED}.
Table \ref{t:Properties} lists properties of \wdj; the cooling age assumes a thin envelope \citep{Fontaine-2001-PASP-CosmoChronology}.

We tested for discrepancies in \hhe\ determined from the ultraviolet and optical spectra by fitting the hydrogen abundance with each wavelength range individually, with \Teff\ and \logg\ fixed to the values in Table~\ref{t:Properties}, and at the $\pm1\sigma$ values. 
The best-fitting values of \hhe\ agreed within $2 \sigma$; we conclude that there is no significant discrepancy between the hydrogen abundance measured from the optical and ultraviolet data.

\begin{table}
\centering
\caption{Stellar properties of \wdj.}
\label{t:Properties}
\begin{tabular}{ l Z }
\hline
Parameter & \textrm{Value} \\
 \hline
 $\varpi$ & 10.17 \pm 0.08 \,\mathrm{m\,arcsec}\\
 $\dpi$ & 98.3 \pm 0.8 \,\mathrm{pc}\\
 $\Teff$ & 17\,970 ^{+140}_{-170} \,\mathrm{K}\\
 $\logg$ & 8.04 \pm 0.05\\
 \hhe & -1.08 \pm 0.08 \\
 $\log q\,^{a}$ & -8.92\\
 $\mathrm{t}_\mathrm{cool}$ & 125\,\mathrm{Myr}\\
 M$_\mathrm{WD}$ & 0.617 \,\mathrm{\Msun}\\
 R$_\mathrm{WD}$ & 0.0131 \,\mathrm{\Rsun}\\
 \hline
\end{tabular}
\newline
\begin{flushleft}
{$^a$\,$\log q$\,=\,$\log$($M_{\mathrm{cvz}}\,/\,M_{\ast}$); mass fraction of the convection zone}
\end{flushleft}
\end{table}

We briefly explored two phenomena that are known to affect ultraviolet spectra of white dwarfs.
\citet{Gaensicke-2018-MNRAS-BroadeningLya} have shown that broadening of Ly~$\alpha$ by neutral helium depends strongly on both \Teff\ and \hhe, and can result in asymmetric Ly~$\alpha$ line profiles.
As the fit of the blue wing of the Ly~$\alpha$ ($1140$~--~$1200$\,\AA) is noticeably poor in our analysis we investigated this feature with a small grid of models, including this broadening feature, around the best-fitting parameters.
We fitted \Teff\ and \hhe\ for three fixed values of \logg\ ($8.00,\,8.04,\,8.08$: the canonical, spectroscopic fit, and photometric fit values),  obtaining results that did not differ significantly from our determined values.
From a visual inspection these improved the fit in the blue wing of Ly~$\alpha$, but overall produced a  worse FUV model.
Therefore, we excluded this feature in our best-fitting model.

Pure hydrogen atmosphere white dwarfs with  $\Teff\lesssim20\,000$\,K exhibit a broad absorption feature near $1400$\,\AA, which grows in strength as \Teff\ decreases \citep{Greenstein-1980-ApJ-Quasimolec, Wegner-1982-ApJL-Quasimolec, Wegner-1984-AJ-Quasimolec}.
This has been identified as Ly~$\alpha$ absorption caused by the H$_2^+$ quasi-molecule \citep{Koester-1985-AAP-Quasi-molecular, Nelan-1985-ApJL-Quasimolec}.
A visual inspection suggested that this feature is not present in our COS spectrum.
Nonetheless, given the unusually high abundance of hydrogen, we experimented with including this feature in our models, but did not achieve an improved fit.
We conclude that, despite the ultraviolet and optical spectra being dominated by hydrogen, the H$_2^+$ feature does not form in the helium-dominated atmosphere of \wdj.

\subsection{Metal abundances}
\label{s:ZAbund}

\begin{table}
\centering
\caption{Wavelengths of absorption lines used to determine the relative abundances of each element. Those in italics are resonant ISM absorption lines. }
\label{t:Linelist}
\begin{tabular}{ l Z }
\hline
Ion & \textrm{Vacuum\ Wavelengths\ }($\AA$) \\
\hline
\Ion{C}{i}   & 1140.35, 1277.55, 1328.83, 1329.09, 1329.10,\\
             & 1329.58, 1329.60 \\
\Ion{C}{ii}  & 1323.91, 1323.95, \emph{1334.53}, 1335.66, \emph{1335.71} \\
\Ion{O}{i}   & 1152.15, \emph{1302.17}, 1304.86, 1306.03, 7774.08,\\
             & 7776.30, 7777.53, 8448.68 \\
\Ion{N}{i}   & \emph{1134.17}, \emph{1134.41}, \emph{1134.98}, 1167.45, 1168.54 \\
             & 1176.51, \emph{1199.55}, \emph{1200.22}, \emph{1200.71}, 1243.18 \\
             & 1243.31, 1310.54, 1319.68, 1411.95 \\
\Ion{Mg}{ii} & 4482.38, 4482.58, 7898.54, 9220.78 \\
\Ion{Al}{ii} & 1189.19, 1190.05, 1191.81, 3587.58, 3901.78,\\
             & 4664.35, 5594.85, 7044.03,\\
             & 7058.66, 7473.47 \\
\Ion{Al}{iii}& 1379.67, 1384.13 \\
\Ion{Si}{ii} & \emph{1190.42}, \emph{1193.29}, 1194.50, 1197.39, 1246.74,\\
             & 1248.43, 1250.09, 1250.44, 1251.16, \emph{1260.42},\\
             & 1264.74, 1265.00, \emph{1304.37}, 1305.59, 1309.28,\\
             & 1309.45, 1346.88, 1348.54, 1350.07, 1350.52,\\
             & 1350.66, 1352.64, 1353.72, 3857.11, 4129.22,\\
             & 4132.06, 5057.39, 6348.86, 6373.13\\
\Ion{Si}{iii}& 1108.36, 1109.94, 1109.97, 1113.17, 1113.20,\\
             & 1113.23, 1294.55, 1296.73, 1298.89, 1298.95,\\
             & 1301.15, 1303.32, 1312.59, 1417.24 \\
\Ion{P}{ii}  & 1149.96, 1152.82, 1154.00, 1155.01, 1249.83\\
\Ion{S}{i}   & 1316.54, 1425.03, 1425.19\\
\Ion{S}{ii}  & 1124.40, 1124.99, 1131.06, 1131.66, \emph{1250.58},\\
             & \emph{1253.81}, \emph{1259.52}\\
\Ion{Ca}{ii} & 3180.25, 3737.97, 3934.78, 3969.59, 8544.44\\
\Ion{Fe}{ii} & 1126.42, 1126.59, 1126.88, 1142.31, 1142.36,\\
             & 1144.94, 1147.41, 1151.16, 1262.14, 1266.53,\\
             & 1266.68, 1267.42, 1275.14, 1275.78, 1311.06,\\
             & 1359.06, 1361.37, 1364.59, 1364.76, 1371.02,\\
             & 1374.83, 1374.94, 1375.17, 1379.47, 1392.82,\\
             & 1408.48, 3228.67, 5019.84, 5170.47, \\
\Ion{Fe}{iii}& 1122.52, 1124.87, 1126.73, 1128.04, 1128.72,\\
             & 1130.40, 1131.19, 1386.15\\
\Ion{Ni}{ii} & 1119.33, 1133.73, 1137.09, 1139.64, 1154.42,\\
             & 1164.28, 1164.58, 1171.29, 1173.30, 1317.22,\\
             & 1335.20, 1345.88, 1370.13, 1374.07, 1381.29,\\
             & 1381.69, 1393.32\\
\hline
\end{tabular}
\end{table}

We identified photospheric absorption lines from nine metals: carbon, oxygen, magnesium, silicon, phosphorus, sulphur, calcium, iron, and nickel (wavelengths in Table~\ref{t:Linelist}).
We fitted these lines by iteratively adjusting the abundances until a visual best fit was found for both the ultraviolet and optical spectra.
Uncertainty estimates are made from the spread of individual line strengths and the change in abundance required after a $\pm 1\sigma$ adjustment in \Teff\ and/or \logg.
We also determine upper limits on the abundances of aluminium and nitrogen (Table~\ref{t:Abundances}).

In order to fit the narrow absorption features accurately, synthetic spectra must be convolved with an empirical line profile that reproduces instrumental broadening of narrow lines.
Owing to the gradual degradation of the COS instrument, the line spread function (LSF) has become non-uniform and non-Gaussian.
Characterisations of these functions are available online\footnote{\url{www.stsci.edu/hst/cos/performance/spectral_resolution/}}.
We interpolate a LSF for the wavelengths corresponding to each pixel and convolve these with the model spectra.
This allows a better fit to absorption line profiles, particularly where adjacent lines are blended together.
Absorption lines in the X-shooter spectra were convolved with a uniform Gaussian profile based on the instrumental characteristics\footnote{\url{www.eso.org/sci/facilities/paranal/instruments.html}}.

Close inspection of Fig.~\ref{f:MetalAbsorption} \& \ref{f:MetalAbsorptionOptical} reveals that not all absorption lines are well reproduced by the best-fitting model.
For example, although the broader absorption features at $\simeq 1260$\,\AA\ are well fit, some narrower \Ion{Si}{ii} lines in the ultraviolet are too strong in the model, while some optical absorption features appear slightly too weak.
Some lines are blended with resonant absorption from the ISM \citep{Mauche-1988-ApJ-ISMAbs}.
This variable quality of fit~--~in particular for iron lines~--~reflects the varying accuracy and completeness in measurements of atomic data which confirm or replace theoretical calculations.
These data are essential to much of the underlying work in astrophysics and \emph{must be supported} by the astronomical community \citep{Nave-2019-BAAS-AtomicDataNeed}.

\begin{table}
\centering
\label{t:Abundances}
\begin{tabular}{ l X X C C }
\hline
 Element & \log\mathrm{[Z/He]} & \tau_{\mathrm{diff}} & \textnormal{Diffusion Flux} & X_\mathrm{cvz} \\
 & & [10^3\,\mathrm{yr}] & [10^6\,\mathrm{g\,s^{-1}}] & [10^{18}\mathrm{g}] \\
 \hline
 C & -\,6.1  \pm 0.1 & 8.6 & 13 & 3.4 \\
 N & <-7.0 & 7.7 & - & - \\
 O & -\,4.8  \pm 0.1 & 7.0 & 410 & 91.6\\
 Mg & -\,5.6 \pm 0.1 & 5.1 & 140 & 22.0 \\
 Al & <-\,6.5 & 4.7 & - & - \\
 Si & -\,5.6 \pm 0.1 & 4.7 & 170 & 25.5 \\
 P & -\,8.1 \pm 0.1 & 4.2 & 1.1 & 0.1 \\
 S & -\,6.6 \pm 0.1 & 4.2 & 22 & 2.9 \\
 Ca & -\,6.9 \pm 0.1 & 3.5 & 16 & 1.8 \\
 Fe & -\,6.4 \pm 0.2 & 2.6 & 98 & 8.0 \\
 Ni & -\,7.4 \pm 0.1 & 2.5 & 11 & 0.8 \\
 \hline
 Total & & & 881 & 157\\
 \hline
\end{tabular}
\end{table}

\section{Composition of Accreted Material}
\label{s:Comp}

The abundances of metals in a white dwarf's photosphere are a result of the composition of the parent body, the duration of the accretion event, and the rate at which each element gravitationally settles below the photosphere. \citet[eq.~4]{Koester-2009-AA-NewAccDiff} described the time-dependent convection zone mass $X_\mathrm{cvz}$ of a particular element, $Z$, which depends on each element's diffusion timescale, $\tau_{\mathrm{diff}}$.
The relative accretion rate of each metal $\dot{M}(Z, t)$ is assumed to be proportional to its relative abundance in the parent body.
Thus, with measured photospheric abundances and computed diffusion timescales, one can determine the bulk composition of the parent body.

It is important to note that 3D atmospheric models are needed to truly understand the mixing processes in white dwarfs; \citet{Cukanovaite-2018-MNRAS-3DDB, Cukanovaite-2019-MNRAS-3DHyHe} have made significant progress, but are yet to extend to atmospheric compositions like that of \wdj.
Similarly, \citet[]{Bauer-2018-APJ-ThermohalineMixing} showed that thermohaline mixing in hydrogen atmospheres can change the accretion rates by orders of magnitude, and to some extent the inferred compositions, but no equivalent study has been published for helium atmospheres.
Thus, for the remainder of this work, we take the conventional approach based on static, one-dimensional envelope models with homogeneous abundance profiles:
\citet{Koester-2009-AA-NewAccDiff} examined a simplified accretion episode wherein the accretion rate is described by a step function; switching on at $t=0$, and then falling instantaneously to zero after some finite duration.
This results in three key phases of an accretion episode, for which we adopt the clearer terminology from \citet[]{Swan-2019-MNRAS-Diversity}:
\begin{itemize}
\item \emph{Increasing Phase:} when accretion begins, the rate of gravitational settling is initially negligible.
Metal abundances in the atmosphere increase linearly with time, and abundance ratios are equal to those of the parent body.
This strictly holds only while $t \ll \tau_\mathrm{diff}$, but is a reasonable approximation up to $t < \tau_\mathrm{diff}$.
\item \emph{Steady State:} after $\simeq 5$ diffusion timescales have passed, accretion and diffusion reach a balanced state where atmospheric abundances are constant. Relative abundances must be multiplied by the ratio of element diffusion timescales to recover the ratios in the parent body.
This phase lasts for as long as accretion is ongoing.
\item \emph{Decreasing Phase:} metal abundances in the photosphere exponentially decay on their individual diffusion timescales.
Atmospheric abundance ratios diverge from those of the parent body.
Absorption lines can remain detectable for several diffusion timescales after accretion stops.
\end{itemize}

\begin{figure*}
\includegraphics[width=\textwidth]{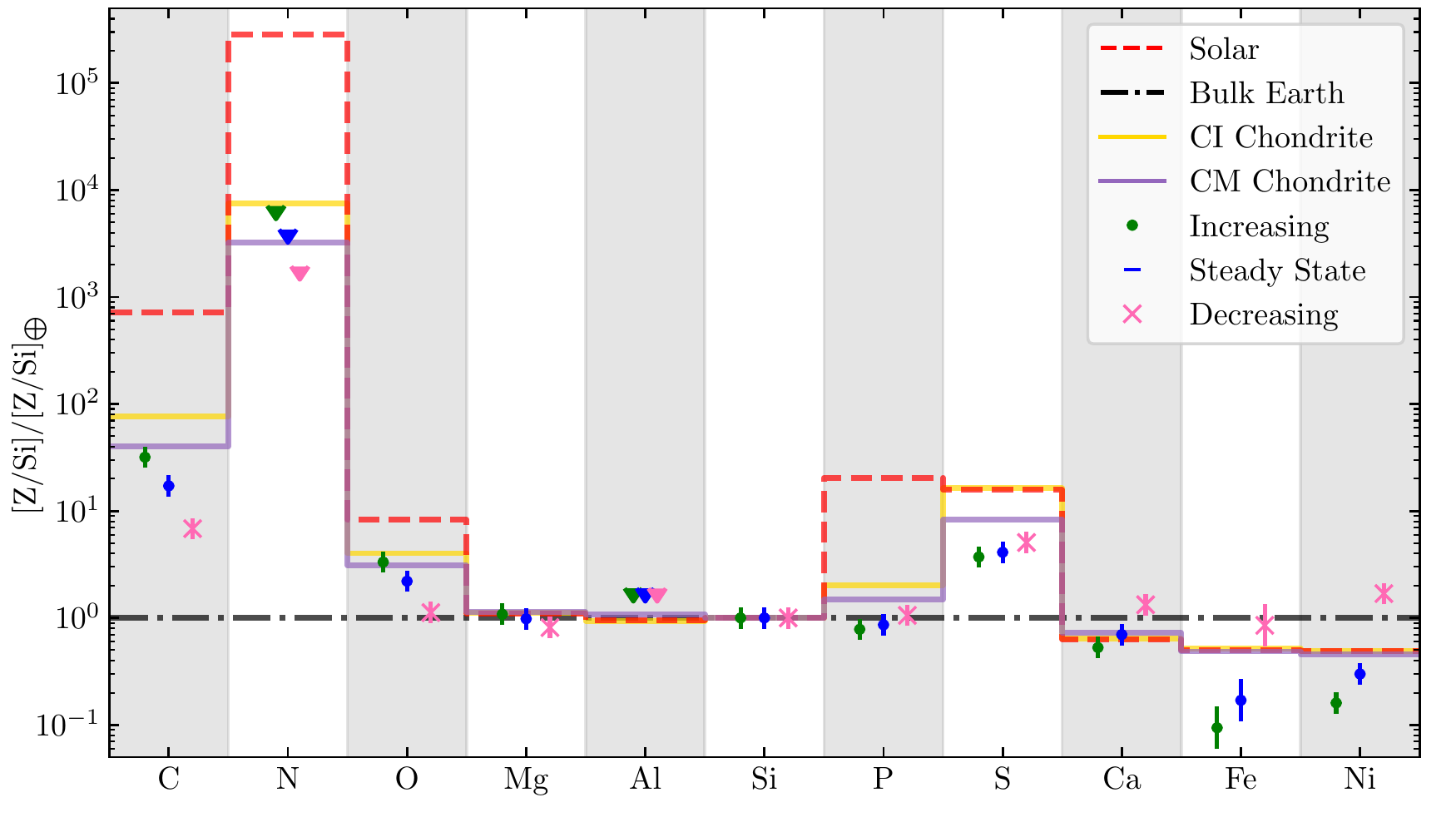}
\caption{Abundances of metals, relative to silicon, in the parent body, normalised to the corresponding bulk Earth ratio, and calculated for each accretion phase.
The decreasing phase is shown for 2$\tau_\mathrm{diff}(\mathrm{Si})$ after accretion ends.
Chondritic compositions show the closest match to the abundances of most of the measured elements; carbon, oxygen and sulphur are best matched to CM chondrite composition. 
Upper limits are shown by triangles.}
\label{f:ParentComp}
\end{figure*}

Using the photospheric metal abundances and diffusion timescales listed in Table~\ref{t:Abundances}, we calculated the parent body abundances~--~relative to silicon~--~for each phase.
For the decreasing phase, the abundances depend on the assumed interval since the abrupt end of accretion; we adopt 2$\tau_{\mathrm{diff}}(\mathrm{Si})$ (9320)\,yr.
These abundances, ordered by atomic mass, are shown in Fig.~\ref{f:ParentComp}, shown with elements ordered by condensation temperature.

The increasing and steady state abundances are similar; both are rich in carbon, sulphur, and oxygen, but poor in iron and nickel, relative to the bulk Earth.
The abundances suggest \wdj\ is accreting the debris of a rocky planetesimal broadly similar to carbonaceous chondrite (C chondrite) material \citep{Kallemeyn-1981-gca-CChondrites, Wasson-1988-PTRS-Chondrites, Lodders-2003-ApJ-Abundances,Lodders-2010-ASSP-CIChond}, most well matched to the CM chondrite subgroup\footnote{CM chondrites are the most commonly found type of C chondrites, named after the Mighei meteorite. Subclasses of C chondrites are distinguished by chemical and petrological properties.}.
Iron and nickel abundances are both well below bulk Earth levels, and slightly below the two illustrated classes of C chondrite.
Within the uncertainties of our measurements, and the diversity among Solar-system chondrites, the abundances of the planetesimal accreted by \wdj\ suggest that it had not undergone major alterations in the form of melting or differentiation.

The increasing phase can last for $\sim\tau_{\mathrm{diff}}$ (in this case, $\sim10^4$\,yr), whereas the steady state phase is estimated to last for $10^{5.6\pm1.1}$\,yr, \citep[based on estimated lifetimes of debris discs]{Girven-2012-ApJ-DiskLifetimes}.
The increasing phase does not produce a significantly favourable composition - the abundances of iron and nickel are better reproduced by the steady state phase.
Thus, we favour the steady state phase composition.

Adopting that \wdj\ is in the decreasing phase brings the iron and nickel abundances in line with those of the bulk Earth, but leaves the volatiles carbon and sulphur highly over-abundant.
\citet{Wilson-2016-MNRAS-CORatio} have shown that the C/O abundance ratio can distinguish between C chondrite and bulk Earth compositions (for which $\log\mathrm{[C/O]}\simeq -1$ \citep{Wasson-1988-PTRS-Chondrites, Lodders-2010-ASSP-CIChond}, and $-1.8\,$--$\,-2.7$ \citep{Allegre-2001-EPSL-BulkEarth}, respectively).
The C/O ratio is not strongly affected by the assumptions about the accretion history, as both elements have relatively similar sinking times (Table\,\ref{t:Abundances}):
We calculate $\log\mathrm{[C/O]}=-1.30, -1.39$, and $-1.50$ for the increasing phase, steady state, and decreasing phase, respectively, with uncertainties of $\pm0.14$.
Hence, the C/O ratio of \wdj\ alone indicates that the accreted material is similar to Solar System C chondrites, and enhanced compared to bulk Earth.
We conclude that the abundances implied by assuming \wdj\ is in the decreasing phase do not match bulk Earth, or other Solar System compositions known to us.

The convection zone of \wdj\ has a mass fraction of $\log q = \log(M_{\mathrm{cvz}}\,/\,M_{\ast}) = -8.92$ \citep[based on][]{Koester-2009-AA-NewAccDiff}.
From the measured abundances, we determine $1.57\times10^{20}$\,g of metals in the convection zone.
A spherical object of this mass, and a density of $3$\,g\,cm$^{-3}$, would have a radius of $\simeq23$\,km; slightly smaller than Mathilde~--~incidentally a C-type asteroid rich in carbon \citep{Yeomans-1997-Science-Mathilde}.
This is a lower limit on the size of the object.
For the steady state phase to maintain this mass requires a diffusion flux of $\Maccrate$, which is by definition equal to the accretion rate $\dot{M}_\mathrm{acc}$.
This value is typical of the measured accretion rates for polluted white dwarfs of similar temperatures, although helium-dominated white dwarfs show a large scatter, extending to higher accretion rates \citep[see also fig. 4. of\ \citealt{Bauer-2018-APJ-ThermohalineMixing}]{Koester-2006-AA-AccDiff, Farihi-2012-MNRAS-AccretionEpisodes}.

The water content of the accreted material can be estimated by calculating the oxygen budget \citep[first explored by][]{Klein-2010-ApJ-MinorPlanet}; comparing the abundance of oxygen with the abundances of elements that are in common minerals.
The assumption that the geochemistry of extra-solar rocky bodies resembles that of the Solar System was demonstrated by \citet{Doyle-2019-Science-OFugacity}.
In the bulk Earth, almost all oxygen is in the metal-oxides common to rocky material; principally SiO$_2$, MgO and FeO, with smaller contributions from other minerals (e.g. Al$_2$O$_3$, CaO), and a negligible amount of water.
In planetesimals formed beyond the snowline, a large fraction of the oxygen budget comes from water.

\begin{figure*}
\includegraphics[width=\textwidth]{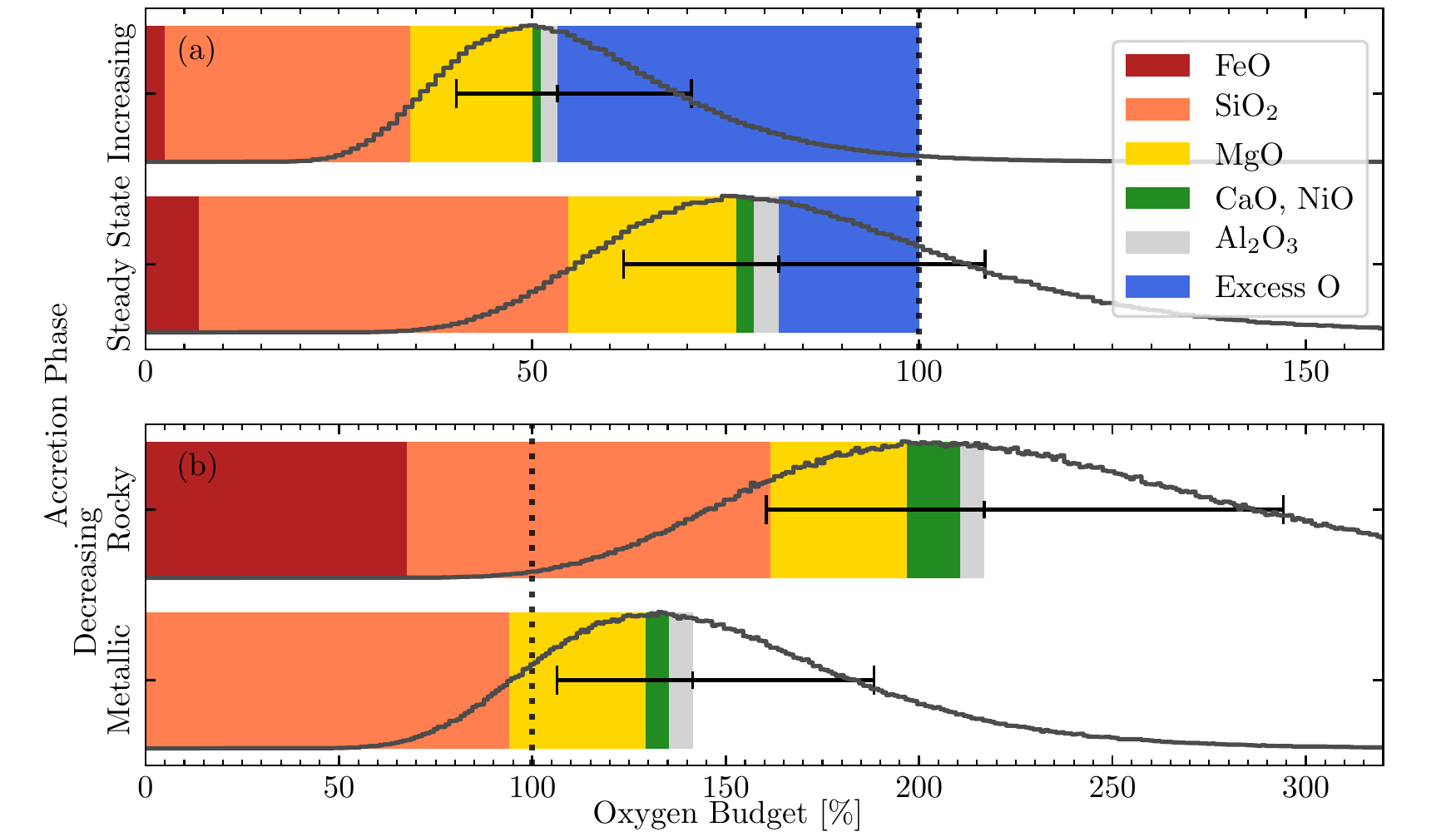}
\caption{The results of oxygen budget simulations for (a) increasing and steady state phases; (b) decreasing phase, $2\tau_\mathrm{diff}(\mathrm{Si})$ after accretion ends, for a completely rocky parent body and for one with metallic iron and nickel.
NB: Axis scales differ between (a) and (b).
Solid lines show the distribution of oxygen budget usage in our simulations, for each phase: 
A higher percentage indicates a relative paucity of oxygen compared to the abundances of mineral-forming elements.
The error bars mark the $50 \pm 34.1$ percentiles of the distributions.
The coloured bars show the breakdown of the median oxygen budget by minerals.}
\label{f:OxygenBudget}
\end{figure*}

Using the compositions determined for each accretion phase, we examine the oxygen budget in the accreted material.
We use Monte Carlo methods to sample the number of atoms of each mineral forming element in our simulated composition. These are multiplied by the empirical ratio for the chemical compound ($\frac{y}{x}$ for Z$_x$O$_y$; e.g. $\frac{2}{1}$ for SiO$_{2}$), summed, and then compared to the total number of oxygen atoms in the simulation.
For rocky material devoid of water, these numbers should be equal; \textit{i.e.} 100~per~cent of the oxygen budget is used.
A median less than 100~per~cent indicates an oxygen excess, best explained by water in the parent body.
If the median is beyond 100 per~cent, some fraction of the parent body would be metallic.
This is not uncommon for iron and nickel, but a metallic component including other mineral-forming elements is unlike any known Solar System body.
\citet{Doyle-2019-Science-OFugacity} have shown that the chemistry resulting from rock formation in exo-planetary systems is broadly similar to that of the Solar System.

As we only have an upper limit on the abundance of aluminum, $\log\mathrm{[Al/He]}<-6.5$, we included it in the analysis at its expected abundance for bulk Earth and C chondrite compositions; $\log\mathrm{[Al/Si]}=-1.05$ \citep[e.g.][]{Lodders-2010-ASSP-CIChond}, but with zero uncertainty.
Not accounting for aluminium would cause a significant systematic offset in the oxygen budget, and our adopted approach is robust as the abundance of aluminium does not vary significantly between bulk Earth and chondritic compositions.

The resulting distributions are approximately lognormal.
In the increasing and steady state phases (Fig.~\ref{f:OxygenBudget}a) -- we find an oxygen excess, $\iota$, of $46^{+13}_{-17}$ and $16^{+20}_{-27}$ per~cent, respectively (uncertainties are from the $16^{\rm th}$ and $84^{\rm th}$ percentiles of the distributions).
The corresponding median mass fractions of water in the accreted material are $29$ and $8$ per~cent.
C chondrite meteorite samples typically contain $10$\,--\,$25$ per~cent water by mass, varying slightly between subgroups \citep{Garenne-2014-GCA-Chondrites}.
We find excess oxygen in 98 and 74 per~cent of simulated compositions, respectively.

\citet[]{Jura-2010-AJ-WaterSurvival} suggested that survival of water through the AGB phase is limited to that sequestered tens of kilometres within a rocky body.
For example, \citet[]{Malamud-2016-ApJ-PMSIce1, Malamud-2017-ApJ-PMSIce2, Malamud-2017-ApJ-PMSIce3} predict that a $\simeq50$\,km minor planet formed at $\simeq40$\,AU around a $2$\,\Msun\ progenitor (evolving into a $0.59$\,\Msun\ white dwarf) would retain half of its water through stellar evolution.
The eccentric Kozai-Lidov mechanism has been shown to cause instabilities in wide orbit planetesimals large enough to cause their accretion \citep{Stephan-2017-ApJL-Icebergs}.

For the decreasing phase (Fig.~\ref{f:OxygenBudget}b) we show the oxygen budget at $2\tau_\mathrm{diff}(\mathrm{Si})$ after accretion ends, for two different mineralogies;
\emph{rocky}, in which all rock-forming elements are oxidised, and \emph{metallic}, where iron and nickel are in their metallic form and do not affect the oxygen budget.
Other common oxidation states (e.g. Fe$_2$O$_3$ and NiO$_2$) for iron and nickel are not investigated.
The oxidation state of iron has been shown to vary across subgroups of C chondrites, but always some iron is in an oxidised state \citep{Garenne-2019-MPS-IronChondrites}.
The oxidation state of the parent body accreted by \wdj\ may lie between our simplified rocky and metallic cases.
For both cases, the result is an oxygen deficit.

We calculated a parent body composition, and subsequent oxygen budget, for a range of times after the end of accretion, to find when the computed abundances become unrealistic.
We accounted for wholly rocky mineralogy, and for a parent body with purely metallic iron and nickel.
The constraints on the maximum duration of the decreasing phase are given in Table~\ref{t:ObudgetLimits}.
These are the times at which the computed composition distributions indicate an oxygen deficit at the 95$^\textrm{th}$ and 99.7$^\textrm{th}$ percentiles (two and three sigma confidence).
These upper limits are nearly two orders of magnitude shorter than estimated disc lifetimes of $10^{5.6\pm1.1}$\,yr \citep{Girven-2012-ApJ-DiskLifetimes}, although within two sigma.
We interpret this, in combination with the C/O ratio mentioned above, as evidence that the decreasing phase scenario is unlikely to be correct for \wdj.

\begin{table}
\centering
\caption{Time constraints on the decreasing phase for different mineralogies ($\tau_{\mathrm{Si}} =  4660$\,yr). }
\label{t:ObudgetLimits}
\begin{tabular}{ l X X }
\hline
 & \textrm{95}^{th}\textrm{percentile} & \textrm{99.7}^{th}\textrm{percentile} \\
 & \multicolumn{2}{X}{[\tau_{\mathrm{diff}}(\mathrm{Si})]} \\
 \hline
 Metallic Fe \& Ni & 2.32 & 3.23 \\
 FeO \& NiO &  1.40 & 1.78 \\
 \hline
\end{tabular}
\end{table}

We next consider the possible mass of water delivered to \wdj, $M_{\mathrm{H_{\mathrm{2}}O}}$, adopting the steady state phase and the median oxygen excess value.
\textit{Assuming at least five diffusion timescales have passed} since the start of the accretion episode (in order to reach the steady state), we derive 
    $M_{\mathrm{H_{\mathrm{2}}O}}\,\geq\,5\tau_{\mathrm{diff}}(\mathrm{Si})\,\times\,\dot{M}_{\mathrm{acc}}\,\times\,\iota = 5\times10^{19}\,\mathrm{g}$.
This is a small amount compared to that held by some Solar System objects~--~four orders of magnitude less than that in Ceres ($\sim10^{23}$\,g).
We discuss the possible implications of water accretion, and the total hydrogen content in the atmosphere of \wdj, further in Section~\ref{s:HinHe}.

\section{No Infrared Excess}
\label{s:IR}

Coverage of infrared wavelengths is provided by archival \textit{Spitzer} data (Fig.~\ref{f:SED}), and the Near Infrared (NIR) X-shooter spectrum.
The \textit{Spitzer} photometric data lie within $\pm1\sigma$ of synthetic photometry for the featureless blackbody of the same temperature.
The NIR spectrum is consistent with this evidence, and Fig.\,\ref{f:SED} shows it to be compatible with the Rayleigh-Jeans tail of the atmospheric model.

As we find a steady state accretion phase the most likely, the presence of an accretion disc is implied~--~but is not confirmed by these data.
Among single white dwarfs, metal pollution is detected far more frequently than warm debris discs  \citep{Wilson-2019-MNRAS-UV-IR-Study} .
Disc detections correlate with high accretion rates, typically $\dot{M}_{\mathrm{acc}} \gtrsim 10^9$\,g\,s$^{-1}$ \citep{Bergfors-2014-MNRAS-FaintDiscs}.
The accretion rate for \wdj\ is slightly lower~--~$\Maccrate$~--~so a non-detection is not unusual for this system; indeed there are multiple examples of white dwarfs that must be accreting material, but have no infrared excess detection.
For example, a thin edge-on disc could be the cause of this non-detection \citep{Bonsor-2017-MNRAS-IRObs}.

\section{Hydrogen in Helium Atmospheres}
\label{s:HinHe}

\wdj\ is the latest addition to the small group 
of peculiar white dwarfs with helium-dominated atmospheres that show anomalously large hydrogen abundances and photospheric metal pollution from accreted planetary material.
These appear to be the extreme outliers from the large population of helium-dominated white dwarfs thought to be accreting hydrogen and metals from water-bearing rocky material \citep{GentileFusillo-2017-MNRAS-He_water}.

\subsection{The argument for water accretion}

\citet[]{Rolland-2018-ApJ-DBAs} measured the hydrogen abundances of a sample of 143 helium rich white dwarfs, the majority of which contain more hydrogen than can be explained by the scenarios they explore.
\citet[]{Rolland-2018-ApJ-DBAs} initially used models that were chemically homogeneous, but \citet{Rolland-2020-arXiv-HDredge} have since implemented more complex abundance profile inputs.
The authors modeled white dwarfs containing a fixed mass of hydrogen, wherein two possibilities arise:
If the initial hydrogen mass is sufficiently large ($\gtrsim 10^{-14}$\Msun), once the white dwarf has cooled to $\simeq25\,000$\,K its envelope becomes stratified as hydrogen floats to the surface, and consequently helium absorption lines are no longer detectable as the photosphere is within the hydrogen layer.
For lower initial hydrogen masses no stratification occurs, but hydrogen is diluted as the convective layer grows deeper, and eventually becomes undetectable.
The hydrogen masses, measured from observations of helium-rich white dwarfs \citep{Koester-2015-AA-DBSurvey, Rolland-2018-ApJ-DBAs}, in the range $12\,000$--$20\,000$\,K cannot have been present throughout the history of those stars, or they would have become stratified and remained so.

Seeking an alternative explanation, \citet[]{Rolland-2018-ApJ-DBAs, Rolland-2020-arXiv-HDredge} examined white dwarfs with negligible initial hydrogen mass, but which accrete hydrogen at a constant rate.
The resulting picture is similar: if this accretion rate is too high ($\gtrsim 10^{-23}$\,\Msun\,yr$^{-1}$ $\simeq 10^3$\,g\,s$^{-1}$), the envelope inevitably becomes stratified as too much hydrogen will be accrued before the convection zone develops, which is subsequently suppressed.
For lower accretion rates, hydrogen remains well mixed and dilutes as the convection zone grows, but the mass of hydrogen accrued cannot produce the observed abundances in the range $12\,000$--$20\,000$\,K.
\citet[]{Rolland-2018-ApJ-DBAs, Rolland-2020-arXiv-HDredge} concluded that the measured hydrogen masses could not be explained by left over hydrogen from the progenitor, nor by a constant rate of accretion.
\citet[]{Rolland-2020-arXiv-HDredge} also sketched out a novel scenario in which hydrogen can be dredged up from deep within the envelope, which has potential to explain a larger portion of the observations.

An accretion scenario will only work if the start of accretion is delayed until when the convection zone begins to grow significantly, at $\simeq24\,000$\,K (fig.~2 of \citealt{Rolland-2020-arXiv-HDredge})
Dynamical simulations have found that planetesimals are not scattered to within the Roche radius of the white dwarf for the first $\simeq 40$\,Myr of its cooling lifetime \citep{Mustill-2018-MNRAS-UnstableWDSystems} - corresponding to $\Teff\simeq23\,000$\,K for a $0.6$\,\Msun white dwarf.
These simulations show that subsequent Roche limit crossings occurred stochastically.
The delay in the onset of accretion demonstrates that the varying accretion history
may provide a natural explanation for the delivery of (water-rich) planetary debris \emph{after} white dwarfs have developed sufficiently deep convection zones that prevent stratified atmospheric configurations.

\subsection{The specific case of WD\hspace{1pt}J2047-1259}

Here, we consider the plausible hydrogen accretion histories for \wdj.
We have found a water mass fraction of approximately eight per~cent in the accreted material from the median oxygen excess in our steady state composition simulations.
Based on the measured abundance of hydrogen and envelope mass from our model, the total mass of hydrogen present in the atmosphere is $M_{\mathrm{H}}~=~3\times10^{22}$\,g ($\sim 10^{-11}$\,\Msun); orders of magnitude above the limits imposed on the initial hydrogen mass $M_{\mathrm{H}}~\lesssim~10^{-14}$\,\Msun\ for the envelope not to become stratified \citep{Rolland-2018-ApJ-DBAs, Rolland-2020-arXiv-HDredge}; according to fig.~4 of \citet[]{Rolland-2020-arXiv-HDredge} a star with this \Teff\ and $M_\mathrm{H}$ should be stratified.

We begin with the case in which a single body delivered all of the accumulated hydrogen during the current accretion episode.
\textit{Assuming no change in the composition or rate of accretion}, the parent body mass required is $\sim10^{24}$\,g; comparable to the mass of Ceres (albeit with a lower water fraction).
The current episode would have to have lasted for approximately $115$\,Myr~--~ most of the cooling age of the star ($125$\,Myr).
Given the stochastic accretion predicted by \citet{Mustill-2018-MNRAS-UnstableWDSystems}, and disc lifetime estimates, which range between $0.01$--$10$\,Myr, \citep{Jura-2008-AJ-Asteroid_pollution, Metzger-2012-MNRAS-RunawayAcc, Girven-2012-ApJ-DiskLifetimes}, the hypothesis of a single parent body dominating accretion onto \wdj\ for $\sim100$\,Myr is unreasonable.
We exclude the possibility that \wdj\ has undergone monotonous accretion from a single object for $>10$\,Myr.
Further, accretion spanning most of the cooling age of the star would require an average rate of $\sim10^{-19}$\,\Msun yr$^{-1}$ to accumulate this mass of hydrogen.
This rate is three orders of magnitude above the limit imposed by \citet[see fig.~8 of the latter]{Rolland-2018-ApJ-DBAs, Rolland-2020-arXiv-HDredge}.

The only plausible explanation involving a single body of this size is that the accretion rate has previously fluctuated to much higher values, such that the duration of the episode could be $\lesssim10$\,Myr.
If the rate had been orders of magnitude higher in the last several diffusion timescales ($\sim10^4$yr) we would see evidence in the abundance ratios. 
The details of the accretion process from circumstellar debris discs are not well understood, but variable infrared emission has been seen in many systems \citep{Xu-2014-ApJL-IRDrop,Xu-2018-ApJ-IRVariable, Farihi-2018-MNRAS-GD56IRVar,Swan-2019-MNRAS-IRvariable, Wang-2019-Nov-Outburst} and calcium emission line strength variations also suggest rapid changes (relative to the disc lifetime) within the gaseous component of debris discs \citep{Wilson-2014-MNRAS-VariableGas, Manswer-2019-Sci-Planetesimal}.
However, these have not been directly linked to changes in the rate of accretion.
To date, there are no observations showing a changing accretion rate at a white dwarf.

Given our understanding of remnant planetary systems, it is almost certain that \wdj\ has accreted planetesimals in the past.
The resulting metal pollution may have long-since sunk out of the photosphere and become undetectable.
However, hydrogen may have been added to the photosphere by any of these past events.
Therefore we consider a scenario wherein the current accretion episode is the latest in a stochastic series over the last $\simeq85$\,Myr (for accretion beginning $\simeq40$\,Myr into the cooling lifetime, as suggested by \citealt{Mustill-2018-MNRAS-UnstableWDSystems}), during which most of the current hydrogen must have been accreted.
This is compatible with fig. 9 of \citet[]{Mustill-2018-MNRAS-UnstableWDSystems} which shows that planetesimal Roche lobe crossings become roughly constant (though still stochastic) at a cooling age of $60$ Myr.
For a canonical white dwarf, $60$\,Myr corresponds to a temperature of $\simeq21\,000$\,K.
According to \citet[fig.~4]{Rolland-2020-arXiv-HDredge}, the maximum permitted hydrogen mass at this temperature, whilst retaining a helium-dominate photosphere, is $\sim10^{-14}$\,\Msun; \emph{three orders of magnitude below the current mass in \wdj}.
We thus conclude that 99.9 per~cent of the hydrogen present in \wdj\ has been delivered by accretion within the last $60$\,Myr.
The large amount of water delivered (compared to other helium atmosphere white dwarfs of similar temperatures) in this system may be attributed to some unusual architecture of the surviving planetary system, or an abundance of water far in excess of the typical planetary systems \citep{Jura-2012-AJ-WaterFractions, GentileFusillo-2017-MNRAS-He_water}.

\section{Conclusions}
\label{s:Conc}

We have characterised of the atmosphere of \wdj\ which contains $1.6\times10^{20}$\,g of metals; carbon, oxygen, magnesium, silicon, phosphorus, sulphur, calcium, iron, and nickel.
We also confirm the previously reported large hydrogen abundance in the helium dominated atmosphere, finding $\Teff\,=\,17,970\pm^{140}_{170}$\,K, $\logg\,=\,8.04\pm0.05$ and $\hhe\,=\,-1.08\pm0.08$.
We show that the photosphere is polluted by rocky material rich in volatile and transitional elements compared to bulk Earth, but consistent with C chondrites (in particular, most similar to the CM chondrites).
It is most likely that accretion is in steady state, and that the parent body probably contained a small amount water; we find excess oxygen in 74~per~cent of simulated steady state compositions, resulting in a median eight per~cent water by mass.

Neither the increasing nor decreasing phase can be definitively excluded, but both are subject to severe constraints to their duration in comparison to the steady state phase.
The increasing phase scenario yields a much higher water fraction by mass in the accreted material (29~per~cent), and a composition that closely resembles C chondrites.
The decreasing phase scenario yields a composition unlike any known planetary body in the Solar System, as well as an implausible oxygen budget.

We infer $3~\times~10^{22}$\,g of hydrogen in the atmosphere of this star.
Although convective mixing processes may have altered the atmosphere of \wdj, no such process can account for its current state.
The star must have acquired its extraordinary abundance of hydrogen within the last $\simeq 60$\,Myr.
We find only two plausible scenarios that could explain this:
\begin{itemize}
\item An ongoing accretion episode of an object with mass similar to (or larger than) Ceres.
The accretion rate was previously significantly higher, in order that the accretion episode last $\le10$\,Myr, but must have been stable at the current level for several diffusion timescales ($\simeq0.01$\,Myr).
\item A stochastic series of accretion episodes, which began within the last $\simeq 85$\,Myr.
Many of these episodes have delivered water-rich material to the white dwarf; within the last $\simeq 60$\,Myr $\sim10^{-11}$\Msun\ of hydrogen was accreted.
This hydrogen has most likely come from a large number of water-bearing planetesimals (e.g chondrites) or from a small number of large objects that contain water.
\end{itemize}
The latter scenario is more consistent with our current understanding of how remnant planetary systems evolve around white dwarfs, and suggests that this exo-planetary system could be either particularly water rich, or currently has an architecture that perturbs an abnormally large number of water-bearing objects on wide orbits.
If all of this hydrogen was delivered as water, it would equate to five per~cent of Earth's oceans.

\section*{Acknowledgements}
MJH is supported by the UK Science and Technology Facilities Council studentship ST/R505195/1, and thanks KMZ and JM for their encouragement during this work.
OT was supported by a Leverhulme Trust Research Project Grant. The research leading to these results has received funding from the European Research Council under the European Union's Seventh Framework Programme (FP/2007-2013) / ERC Grant Agreement n. 320964 (WDTracer).
CJM and BTG were supported by the UK STFC grant ST/P00049.
RR acknowledges funding by the German Science foundation (DFG) through grants HE1356/71-1 and IR190/1-1, and funding from the postdoctoral fellowship programme Beatriu de Pin\'os, funded by the Secretary of Universities and Research (Government of Catalonia) and by the Horizon 2020 programme of research and innovation of the European Union under the Maria Sk\l{}odowska-Curie grant agreement No 801370.
The research leading to these results has received funding from the European Research Council under the European Union's Horizon 2020 research and innovation programme no. 677706 (WD3D).
AS acknowledges support from an STFC studentship.

This work has made use of data from the European Space Agency (ESA) mission {\textit Gaia} (\url{https://www.cosmos.esa.int/gaia}), processed by the {\textit Gaia} Data Processing and Analysis Consortium (DPAC, \url{https://www.cosmos.esa.int/web/gaia/dpac/consortium}).
Funding for the DPAC has been provided by national institutions, in particular institutions participating in the {\it Gaia} Multilateral Agreement.
This work is based on observations collected at the European Southern Observatory under ESO programme 0101.C-0646, and on observations made with the NASA/ESA Hubble Space Telescope, obtained at the Space Telescope Science Institute, which is operated by the Association of Universities for Research in Astronomy, Inc., under NASA contract NAS 5-26555. These observations are associated with programmes \#15073 and \#15474.
This publication makes use of data products from the Two Micron All Sky Survey, which is a joint project of the University of Massachusetts and the Infrared Processing and Analysis Center/California Institute of Technology, funded by the National Aeronautics and Space Administration and the National Science Foundation \nocite{Skrutskie-2006-AJ-2MASS}.
The identification of individual lines made use of the National Institute of Standards and Technology Atomic Spectra Database Lines Form \nocite{Kramida-2018-online-NIST_ASD}.
The Pan-STARRS1 Surveys (PS1) and the PS1 public science archive have been made possible through contributions by the Institute for Astronomy, the University of Hawaii, the Pan-STARRS Project Office, the Max-Planck Society and its participating institutes, the Max Planck Institute for Astronomy, Heidelberg and the Max Planck Institute for Extraterrestrial Physics, Garching, The Johns Hopkins University, Durham University, the University of Edinburgh, the Queen's University Belfast, the Harvard-Smithsonian Center for Astrophysics, the Las Cumbres Observatory Global Telescope Network Incorporated, the National Central University of Taiwan, the Space Telescope Science Institute, the National Aeronautics and Space Administration under Grant No. NNX08AR22G issued through the Planetary Science Division of the NASA Science Mission Directorate, the National Science Foundation Grant No. AST-1238877, the University of Maryland, Eotvos Lorand University (ELTE), the Los Alamos National Laboratory, and the Gordon and Betty Moore Foundation.

\section*{Data Availability}
Data underlying this article are available in the MAST archive at \url{https://mast.stsci.edu/portal/Mashup/Clients/Mast/Portal.html}, and can be accessed with observation IDs ldnu0r010, ldst02010, and ldst02020.

Data underlying this article are available in the ESO Science Archive at \url{http://archive.eso.org/scienceportal/home}, and can be accessed with program ID 0101.C-0646.





\bibliographystyle{mnras}
\bibliography{references}







\bsp	
\label{lastpage}
\end{document}